\documentstyle[12pt]{article}

\newcommand{\nc}{\newcommand}
\nc{\beq}{\begin{equation}}
\nc{\eeq}{\end{equation}}
\nc{\beqa}{\begin{eqnarray}}
\nc{\eeqa}{\end{eqnarray}}

\def\gsim{\mathrel{\rlap{\lower4pt\hbox{\hskip1pt$\sim$}}
    \raise1pt\hbox{$>$}}}       

\input epsf
\newwrite\ffile\global\newcount\figno \global\figno=1

\def\writedef#1{}
\def\figin{\epsfcheck\figin}\def\figins{\epsfcheck\figins}
\def\epsfcheck{\ifx\epsfbox\UnDeFiNeD
\message{(NO epsf.tex, FIGURES WILL BE IGNORED)}
\gdef\figin##1{\vskip2in}\gdef\figins##1{\hskip.5in}
\else\message{(FIGURES WILL BE INCLUDED)}%
\gdef\figin##1{##1}\gdef\figins##1{##1}\fi}
\def\figinsert{}
\def\ifig#1#2#3{\xdef#1{fig.~\the\figno}
\writedef{#1\leftbracket fig.\noexpand~\the\figno}%
\figinsert\figin{\centerline{#3}}\medskip\centerline{\vbox{\baselineskip12pt
\advance\hsize by -1truein\center\footnotesize{  Fig.~\the\figno.} #2}}
\bigskip\endinsert\global\advance\figno by1}
\def\endinsert{}

\begin{document}



\title{\large{\bf Minimum Length from First Principles}}

\author{Xavier~Calmet$^a$\thanks{calmet@physics.unc.edu},
Michael~Graesser$^b$\thanks{graesser@theory.caltech.edu} and
Stephen~D.H~Hsu$^c$\thanks{hsu@duende.uoregon.edu} \\
\\
$^a$Department of Physics and Astronomy, 
UNC Chapel Hill, NC 27599-3255.\\
$^b$California Institute of Technology, Pasadena, CA 91125\\
 $^c$Institute of Theoretical Science, University of Oregon, Eugene OR 97403.\\ }



\maketitle

\begin{picture}(0,0)(0,0)
\end{picture}
\vspace{-45pt}

\begin{abstract} 
We show that no device or gedanken experiment is capable of
measuring a distance less than the 
Planck length. By "measuring a distance
less than the Planck length" we mean, technically, 
resolve the eigenvalues
of the position operator to within that accuracy. 
The only assumptions in
our argument are causality, the 
uncertainty principle from quantum
mechanics and a dynamical criteria for 
gravitational collapse from
classical general relativity called the 
hoop conjecture. The inability
of any gedanken experiment to measure a 
sub-Planckian distance suggests the
existence of a minimal length.
\end{abstract}



\newpage

In this work we show that quantum mechanics and
classical general relativity considered simultaneously imply the existence of a minimal length, i.e.
no operational procedure exists which can measure a
distance less than this fundamental length. The key ingredients used to
reach this conclusion are the uncertainty principle from quantum
mechanics, and gravitational collapse from classical
general relativity.

A dynamical condition for gravitational collapse is given by the hoop
conjecture \cite{hoop}: if an amount of energy
$E$ is confined at any instant to a ball of size $R$, where $R < E$,
then that region will eventually evolve into a black hole\footnote{We
use natural units where $\hbar, c$ and Newton's constant (or $l_P$)
are unity. We also neglect numerical factors of order one.}.

From the hoop conjecture and the uncertainty principle, we immediately deduce the
existence of a  minimum ball of size $l_P$. Consider a particle
of energy $E$ which is not already a black hole. Its size $r$ must
satisfy \beq r \gsim {\rm \bf max} \left[\, 1/E\, ,\,E \, \right]~~,
\eeq where $\lambda_C \sim 1/E$ is its Compton wavelength and $E$
arises from the hoop conjecture. Minimization with respect to $E$
results in $r$ of order unity in Planck units or $r \sim l_P$.
If the particle {\it is} a black hole, then its radius grows with mass: $r \sim E \sim 1/
\lambda_C$. This relationship suggests that an experiment designed (in
the absence of gravity) to measure a short distance $l << l_P$ will
(in the presence of gravity) only be sensitive to distances
$1/l$.

Let us give a concrete model of minimum length. Let the position operator $\hat{x}$ have
discrete eigenvalues $\{ x_i \}$, with the separation between
eigenvalues either of order $l_P$ or
smaller.
(For regularly distributed eigenvalues with
a constant separation, this would be equivalent to a spatial lattice.)
We do not mean to imply that nature implements minimum length in this particular
fashion - most likely, the physical mechanism is more complicated, 
and may involve,
for example, spacetime foam or strings. However, our concrete 
formulation lends itself
to detailed analysis. We show below that this formulation
cannot be excluded by any gedanken experiment, which is strong evidence for the
existence of a minimum length.

Quantization of position does not by itself imply quantization of
momentum. Conversely, a continuous spectrum of momentum does not imply
a continuous spectrum of position. In a formulation of
quantum mechanics on a regular spatial lattice, with spacing $a$
and size $L$, the momentum operator has eigenvalues which are
spaced by $1/L$. In the infinite volume limit the momentum operator can have
continuous eigenvalues even if the spatial lattice spacing is kept
fixed. This means that the displacement operator \beq \label{disp}
\hat{x} (t) - \hat{x} (0) = \hat{p}(0) {t \over M} \eeq does not
necessarily have discrete eigenvalues (the right hand side of
(\ref{disp}) assumes free evolution; we use the Heisenberg picture
throughout). Since the time evolution operator is unitary the
eigenvalues of $\hat{x}(t)$ are the same as $\hat{x}(0)$. Importantly
though, the spectrum of $\hat{x}(0)$ (or $\hat{x}(t)$) is completely
unrelated to the spectrum of the $\hat{p}(0)$, even though they are
related by (\ref{disp}).  A measurement of arbitrarily small displacement
(\ref{disp}) does not exclude our model of minimum length. To
exclude it, one would have to measure a position eigenvalue $x$
{\it and} a nearby eigenvalue $x'$, with $|x - x'| << l_P$.

Many minimum length arguments are obviated by the simple observation of the minimum ball. However,
the existence of a minimum ball does not by itself preclude the
localization of a macroscopic object to very high precision.
Hence, one might attempt to measure the spectrum of $\hat{x}(0)$
through a time of flight experiment in which wavepackets of
primitive probes are bounced off of well-localised macroscopic
objects. Disregarding gravitational effects, the discrete spectrum
of $\hat{x}(0)$ is in principle obtainable this way. But,
detecting the discreteness of $\hat{x}(0)$ requires wavelengths
comparable to the eigenvalue spacing.  For eigenvalue spacing
comparable or smaller than $l_P$, gravitational effects cannot be
ignored, because the process produces minimal balls (black holes)
of size $l_P$ or larger. This suggests a direct measurement of the
position spectrum to accuracy better than $l_P$ is not possible.
The failure here is due to the use of probes with very short wavelength.

A different class of instrument, the interferometer,  is capable of measuring
distances much smaller than the size of any of its sub-components.  Nevertheless, the uncertainty principle and gravitational collapse prevent an arbitrarily accurate measurement of
eigenvalue spacing.  
First, the limit from quantum mechanics. Consider
the Heisenberg operators for position $\hat{x} (t)$ and momentum
$\hat{p} (t)$ and recall the standard inequality \beq \label{UNC}
(\Delta A)^2 (\Delta B)^2 \geq  ~-{1 \over 4} ( \langle [
\hat{A}, \hat{B} ] \rangle )^2 ~~.
\eeq Suppose that the
position of a {\it free} test mass is measured at time $t=0$
 and {\em again} at a later time.
The
position operator at a later time $t$ is \beq \label{P} \hat{x}
(t) = \hat{x} (0) ~+~ \hat{p}(0) \frac{t}{M}~~. \eeq
We assume a free particle Hamiltonian here for simplicity, but the argument can be generalized \cite{PRL}. The commutator between the position operators at $t=0$ and $t$
is \beq [ \hat{x} (0), \hat{x} (t)] ~=~ i {t \over M}~~,
\eeq so using (\ref{UNC}) we have \beq \vert \Delta x (0) \vert
\vert \Delta x(t) \vert \geq \frac{t}{2M}~~.\eeq
We see that at least one of the uncertainties $\Delta x(0)$ or $\Delta x(t)$
must be larger than of order $\sqrt{t/M}$.
As a measurement of the discreteness of $\hat{x}(0)$
requires {\em two} position measurements,
it is limited by the greater of $\Delta x(0)$ or $\Delta x(t)$,
that is, by $\sqrt{t/M}$, 
 \beq \label{SQL} \Delta x \equiv {\rm \bf max}\left[
 \Delta x(0), \Delta x(t) \right]
 \geq
\sqrt{ t \over 2 M }~~, \eeq where $t$ is the time over
which the measurement occurs and $M$ the mass of the object whose
position is measured. In order to push $\Delta x$ below $l_P$, we
take $M$ to be large.  Note that this is not the standard quantum limit \cite{SQL} which can be overcome using refined techniques \cite{Ozawa}.
In order to avoid
gravitational collapse, the size $R$ of our measuring device must
also grow such that $R > M$. However, by causality $R$ cannot
exceed $t$. Any component of the device a distance greater than
$t$ away cannot affect the measurement, hence we should not
consider it part of the device. These considerations can be
summarized in the inequalities \beq \label{CGR} t > R > M
~~.\eeq Combined with (\ref{SQL}), they require $\Delta x
> 1$ in Planck units, or \beq \label{DLP} \Delta x > l_P~. \eeq 

Notice that the considerations leading to (\ref{SQL}), (\ref{CGR})
and (\ref{DLP}) were in no way specific to an interferometer, and
hence are {\it device independent}. We repeat: no device subject
to quantum mechanics, gravity and causality can exclude the quantization
of position on distances less than the Planck length.

It is important to emphasize that we are deducing a minimum length
which is parametrically of order $l_P$, but may be larger or
smaller by a numerical factor.  This point is relevant to the
question of whether an experimenter might be able to transmit the
result of the measurement before the formation of a closed trapped
surface, which prevents the escape of any signal. If we decrease
the minimum length by a numerical factor, the inequality
(\ref{SQL}) requires $M >> R$, so we force the experimenter to
work from deep inside an apparatus which has far exceeded the
criteria for gravitational collapse (i.e., it is much denser than
a black hole of the same size $R$ as the apparatus). For such an
apparatus a horizon will already exist before the measurement
begins. The radius of the horizon, which is of order $M$, is very
large compared to $R$, so that no signal can escape.

An implication of our result is that there may only be a finite number of degrees of freedom per unit volume in our universe - no true continuum of space or time. Equivalently, there is only a finite amount of information or entropy in any finite region our universe.

One of the main problems encountered in the quantization of gravity is a proliferation of divergences coming from short distance fluctuations of the metric (or graviton). However, these divergences might only be artifacts of perturbation theory: minimum length, which is itself a non-perturbative effect, might provide a cutoff which removes the infinities. This conjecture could be verified by lattice simulations of quantum gravity (for example, in the Euclidean path integral formulation), by checking to see if they yield finite results even in the continuum limit.

\bigskip

\noindent
This research supported in part under DOE contracts
DE-FG06-85ER40224, DE-FG03-92ER40701 and DE-FG02-97ER-41036.


\bigskip


\end{document}